# STEP-UP CONVERTER FOR ELECTROMAGNETIC VIBRATIONAL ENERGY SCAVENGER.

*Chitta Saha[1], Terence O'Donnell[1], Jeffrey Godse11[1], Louis Carlioz[1], Ningning Wang[1], Paul McCloskey[1], Steve Beeby[2], John Tudor[2] and Russel Torah[2]*

[1]Tyndall National Institute, Cork, Ireland.
[2]University of Southampton, School of Electronics and Computer Science, Southampton, UK.

## ABSTRACT

This paper introduces a voltage multiplier (VM) circuit which can step up a minimum voltage of 150 mV (peak). The operation and characteristics of this converter circuit are described. The voltage multiplier circuit is also tested with micro and macro scale electromagnetic vibrational generators and the effect of the VM on the optimum load conditions of the electromagnetic generator is presented. The measured results show that 85% efficiency can be achieved from this VM circuit at a power level of 18 μW.

## 1. INTRODUCTION

A significant amount of research has already been done on vibrational power generators using electromagnetic [1-6], piezoelectric [3] [7-9], and electrostatic principles [3] [8]. These generators require the use of a converter circuit to convert the ac-generated voltage to a usable dc level. In particular an electromagnetic generator generally requires a voltage step-up circuit. A suitable voltage step-up circuit for a low voltage energy scavenger has not been previously established and the optimum load conditions of an EM generator with such a converter have not been investigated. There are different topologies which can be used to perform the ac-dc conversion required to convert the low voltage AC generated by an electromagnetic vibrational generator, to a useable DC voltage level. Possible approaches include a transformer followed by a rectifier, a rectifier followed by a DC-DC converter, or a voltage multiplier. For example, Wen J. Li [1] demonstrated a laser micro-machined vibration based power generator with diode based voltage multiplier (VM) circuits which could produce 2V DC, however the optimum conditions for the generator with the multiplier circuit and the efficiency of the circuit were not analyzed. E.P. James [2] developed the prototype EM generator with step-up transformer combined with VM circuits and discussed the circuit efficiency. A. Kasyap [7] presented the piezoceramic composite beam coupled with a flyback converter circuit and also derived the equivalent circuits and verified the optimization theory. S. Roundy [8] described the theoretical analysis of a piezoelectric generator and verified the optimum condition with a resistive load and also demonstrated the generator with a capacitive load rectifier circuit. E. Lefeuvre [9] demonstrated the detailed analysis and optimum conditions of piezoelectric generator with ac-dc converter circuit using synchronized switch damping techniques

Most previous works suggest [4] [5] that micro scale electromagnetic generators typically generate a maximum of 200-500 mV (peak) at 100 Hz frequency, so that step up conversion id required. Size and efficiency are important factors in determining the choice of technique used for the conversion. A relatively large transformer would be required because of the generally low frequencies. The simple full wave direct rectification using diodes, or a voltage multiplier using diodes could not be used because of the minimum diode forward voltage drop of 0.3 V. If the diode in the voltage multiplier circuit is replaced by an active switch (e.g. mosfet or analogue switch) then the voltage multiplier circuit can be used to step up very low generated voltages. However such switches require an additional power supply and hence consume power which affects efficiency.

In this paper we introduce the prototype of a four stage VM circuit, using active switches, which has been built and characterized using as input source a signal generator and the EM vibration harvesting device. The measured and calculated results of the VM circuits for the both inputs are discussed and analyzed.

## 2. VIBRATIONAL GENERATOR.

Two EM generators have been built and tested with the VM circuits. Figure 1 shows the prototype of a micro generator. The macro generator is simply a larger version of this. The detailed structure of the micro generator and the macro generator was explained in the paper [4] and [6] respectively. Table I summarises the main generator parameters.

*Chitta Saha[1], Terence O'Donnell[1], Geffrey Godse11[1], Louis Carlioz[1], Ningning Wang[1], Paul McCloskey[1], Steve Beeby[2], John Tudor[2] and Russell Torah[2]*

*Step-up converter for electromagnetic vibrational energy scavenger.*

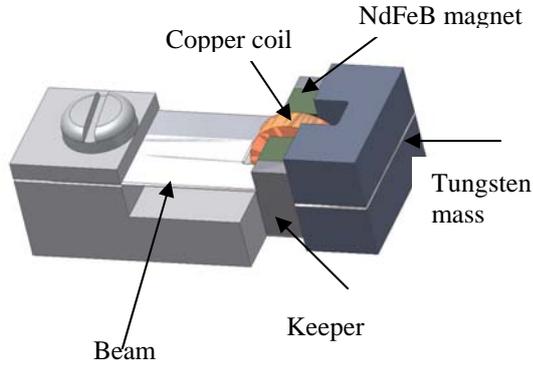

**Figure 1 : Prototype of the EM micro generator.**

**Table 1 : Generator parameters**

| Parameters | Macro generator | Micro generator |
|---|---|---|
| Magnet size (mm) | 15 x 15 x 5 | 2.5x 2 x 1.5 |
| Coil outer diameter (mm) | 19 | 2.4 |
| Coil inner diameter (mm) | 1 | 0.6 |
| Coil thickness (mm) | 6.5 | 0.5 |
| Magnet and coil gap (mm) | 13 | 0.25 |
| Coil turns | 1100 | 2300 |
| Coil resistance (ohm) | 46 | 1613 |
| Acceleration (m/s$^2$) | 0.405 | 0.647 |
| Moving mass (kg) | 0.05 | 0.0066 |
| Resonant frequency | 14.1 | 53 |

The equation of motion of the linear vibrational generator consisting of a mass, m, fixed to a spring with constant k, which is free to move in the x direction is given by [5];

$$m\frac{d^2x}{dt^2} + (D_p + D_e)\frac{dx}{dt} + kx = F\sin\omega t \quad (1)$$

where $D_p$, $D_e$ are parasitic and electromagnetic (EM) damping, and F is the driving force. The displacement at resonance ($\omega=\omega_n$) can be defined as;

$$x = \frac{F\cos\omega t}{[(D_p + D_e)\omega]} \quad (2)$$

The EM damping can be expressed as;

$$D_e = \frac{(N\frac{d\phi}{dx})^2}{R_c + j\omega L + R_l} \quad (3)$$

where d$\phi$/dx, N, $R_c$ and L are the coil flux linkage, number of turns, resistance and inductance respectively and $R_l$ is the load resistance. The maximum electrical power is generated when the EM and parasitic damping are matched, and this can be expressed as;

$$P_{max} = \frac{F^2}{8D_p} = \frac{(ma)^2}{8D_p} \quad (4)$$

However this condition is valid only when the parasitic damping and the resonance frequency are independent of the displacement and consequently the load.
Under these condition the optimum load resistance for the maximum power condition can be defined by;

$$R_{lopt} = \frac{(N\frac{d\phi}{dx})^2}{D_p} - R_c \quad (5)$$

Initially the macro generator and the micro generator were tested by the force control EM shaker with resistive load and then with the VM circuit in order to compare the generated voltage and the maximum power conditions.

**2.1 Generator Characteristics**

Figure 2 and figure 3 show the measured no load peak voltage for the macro and micro generator respectively. The macro-generator shows linear behavior and the micro-generator shows non-linear behavior for this acceleration. This non-linear behavior is caused by a non-linear dependence of the spring constant on the displacement, which can give rise to the discontinuous response as shown. Due to this non-linear effect it was not possible to verify the results from the micro-generator with the linear modeling approach. In order to analyze the non-linear behavior of the mass, damper and spring system, it is necessary to know the maximum linear displacement and the non-linear spring constant [12-13], which is beyond the scope of this work. However the measured results of the macro generator were verified with the linear theoretical model. The graph in figure 4 shows that the maximum power of the macro generator is delivered to the load when the parasitic damping is equal to EM damping which agrees with equation (4) and the maximum power is transferred at 100 ohm load which also agrees with equation (5).

*Chitta Saha[1], Terence O'Donnell[1], Geffrey Godse11[1], Louis Carlioz[1], Ningning Wang[1], Paul McCloskey[1], Steve Beeby[2], John Tudor[2] and Russell Torah[2]*
*Step-up converter for electromagnetic vibrational energy scavenger.*

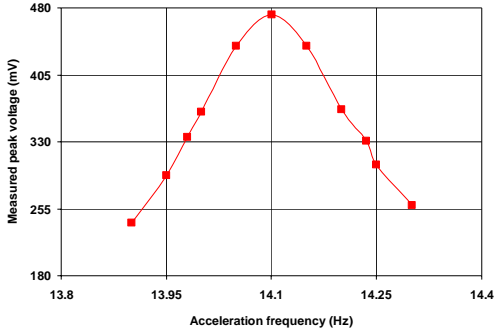

**Figure 2: No-load voltage vs. frequency of macro generator.**

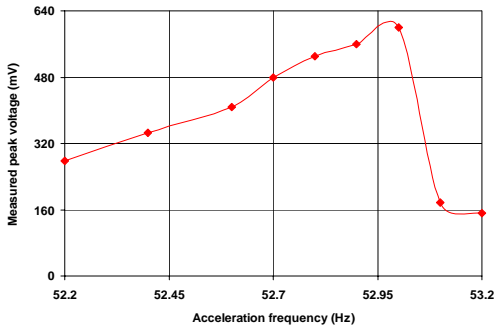

**Figure 3: No-load voltage vs. frequency for micro generator.**

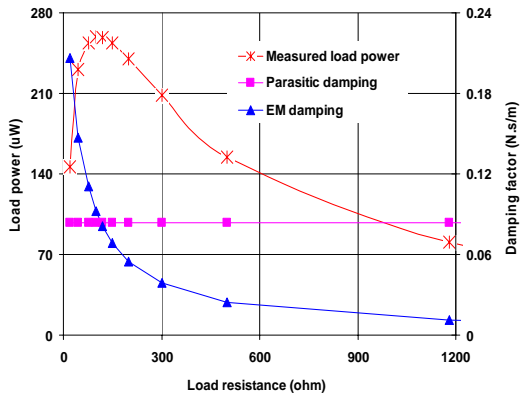

**Figure 4: Measured load power and the calculated damping factor of macro generator.**

The macro-generator has a load voltage of approximately 230 mV (peak) and generates a load power of 260 µW for a 0.4 m/s$^2$ acceleration at 14 Hz frequency. The micro generator generates a load voltage of 350 mV (peak), a load power of 17.5 µW for a 0.65 m/s$^2$ acceleration at 53 Hz frequency.

## 3. ANALYSIS OF THE VM CIRCUIT

It is important to know the fundamental equations of the VM circuits in order to characterize the prototype and analyze the measured results. Figure 5 shows the simple diode capacitor voltage multiplier circuit and its equivalent dc circuit. The output voltage of the capacitor-diode VM is given by [10];

$$V_0 = nV_i - IR_m \qquad (6)$$

where n is the number of stages, $V_i$ is the ac peak generated input voltage, I is the load current and $R_m$ is the resistance of the VM circuit which can be defined for odd and even stages respectively as;

$$R_m = \frac{n(n^2 - 1)}{12Cf} \qquad (7)$$

$$R_m = \frac{n(n^2 + 2)}{12Cf} \qquad (8)$$

where C is the stage capacitor and f is the frequency of the supply voltage.
The voltage transformation factor can be defined by;

$$\lambda = \frac{V_o}{V_i} \qquad (9)$$

Ideally this transformation factor, at no-load, should be equal to the number of stage of the VM circuit. However in a real circuit it is always somewhat less than the number of stages. In this case the efficiency of the circuit is defined by;

$$\eta = \frac{V_0}{nV_i} \qquad (10)$$

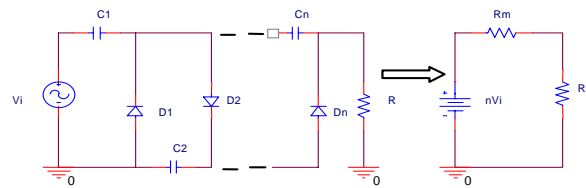

**Figure 5: Equivalent dc circuit of the VM**

The next section gives a brief overview of the prototype of the four stages VM which is built and tested with the signal generator and real vibrational energy harvesting device.

*Chitta Saha[1], Terence O'Donnell[1], Geffrey Godse11[1], Louis Carlioz[1], Ningning Wang[1], Paul McCloskey[1], Steve Beeby[2], John Tudor[2] and Russell Torah[2]*
*Step-up converter for electromagnetic vibrational energy scavenger.*

## 3. PROTOTYPE OF THE VM CIRCUIT

Figure 6 below shows the topology of the four stage VM circuit which is used, where the diodes are replaced by active switches, which are switched on and off using a comparator. Figure 7 shows the prototype of the VM circuit which has been built and characterized. The VM circuit is constructed using Intersil ISL43L120 switches, Seiko S-89530A comparators and Vishay Sprague CTS13107X9010C 100 µF capacitors. The next section describes the measured and calculated results of the VM circuits using a signal generator as input.

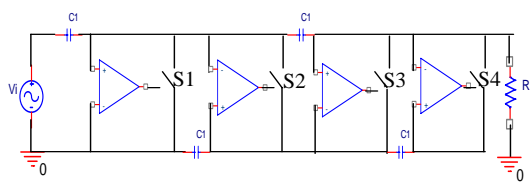

**Figure 6: Topology of four stage VM circuit.**

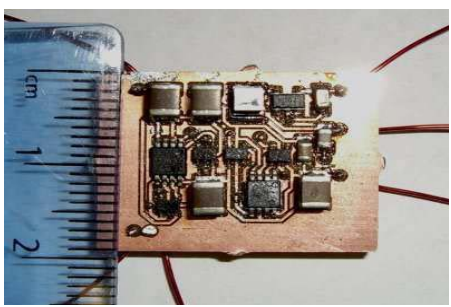

**Figure 7: Prototype of the VM circuit.**

## 4. CHARACTERISTICS OF THE VM

The input and output voltage of the VM circuit are measured using the signal generator as a source in order to characterize the circuit. Figure 8 shows the measured and calculated voltage and the measured and calculated voltage transformation factor for 580 mV (peak) source voltage at 50 Hz frequency. The input voltage and the ratio of the output-input voltage were calculated according to equation (6) and (9) respectively. The measured results agree well with the calculated results. The graph shows that the VM achieves a step-up close to 4 under light loads.

Figure 9 shows the measured load power and efficiency of the VM circuit. This measured results show that 95 % efficiency can be achieved from this VM circuit at maximum load power condition. However these results do not include the power consumption of the switches and the comparators.

In order to determine the power consumption of the switches and comparators the supply current were measured for different loads and for different fixed supply voltages.

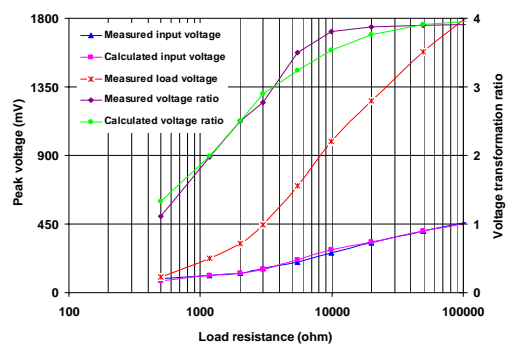

**Figure 8: Measured and calculated voltage and measured and calculated voltage transformation factor.**

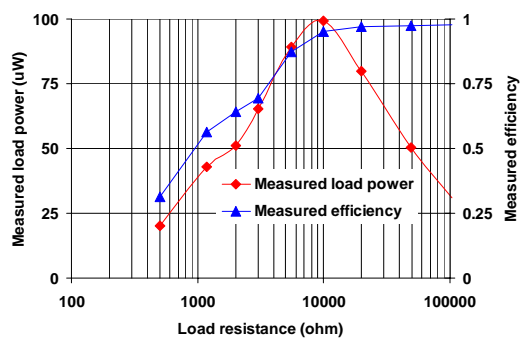

**Figure 9: Measured power and efficiency.**

Figure 10 plots the measured power consumption for a 1.64 V, a 2 V and a 2.4 V supply voltage respectively vs. the load resistance of the converter. We can see from these plots that the power consumption has a dependence on the load and that the power consumption for a 2.4 V supply is considerably higher than for a 1.64 V and a 2V supply. The four comparators were found to dissipate a constant power of approximately 0.5 µW under all load conditions. This closely agrees with the value given in the datasheet of the Seiko S-89530A. Thus it appears that the switches are responsible for the large increase in consumption. If the measured power dissipation of the switches and comparators is included in figure 4 then the overall converter efficiency would be 88 % for 2 V supply voltage.

*Chitta Saha[1], Terence O'Donnell[1], Geffrey Godse11[1], Louis Carlioz[1], Ningning Wang[1], Paul McCloskey[1], Steve Beeby[2], John Tudor[2] and Russell Torah[2]*

*Step-up converter for electromagnetic vibrational energy scavenger.*

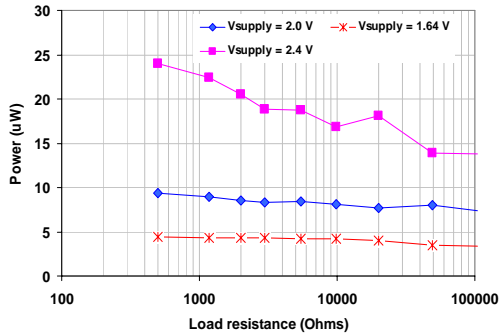

**Figure 10: Power dissipation of the switches and comparators for different supply voltages.**

The next section will examine the optimum conditions of the generator with the VM circuits and the efficiency of the VM circuit with real EM energy harvesting device.

### 6. CHARACTERISTIC OF THE VM WITH VIBRATION GENERATOR.

Figure 11 and figure 12 shows the generated voltage, the load voltage and the voltage transformation factor vs. load resistance where the VM circuit is supplied by the vibration generators. The vibration generators are excited at their resonance frequency with the acceleration levels given in table I. The coil resistance and the resistance of the VM circuit according to equation (8) for macro generator are 46 ohm and 4.3 k and for micro generator are 1.6k and 1.2 k respectively. The generated voltage and the voltage transformation factor are calculated from the equation (6) and (9) respectively. The measured results agree well with the calculated values. However for less than 2 k load for the macro generator and for less than 10 k load resistance for the micro generator, the measured voltage and transformation factor did not match with the calculated values due to very low transformation factor. In this region most the generated voltage is dropped across the coil resistance and the internal VM circuit impedance. Figure 13 and figure 14 show the comparison of the load power of the macro and the micro generator with resistive load and with the VM circuits. This measured results show that the 80-85 % efficiency is achieved of the VM circuit without considering switches and comparators loss. It can be seen that the optimum load resistance required to achieve maximum load power changes significantly when the converter circuit is used. We can see from the power graphs that for the macro and micro generator, the maximum power is delivered at 5.5 k Ω and 50 kΩ load when the VM is attached with a step-up ratio of 3.2 and 3.8 respectively. This compares with optimum load resistances of 100 Ω and 3 kΩ for the generator attached directly to a load. Because the VM circuit steps up the voltage, it also performs an impedance transformation of $\lambda^2$ on the load impedance seen by the generator. Thus with an ideal step-up ratio of 4, the impedance seen by the generator is $1/16^{th}$ the actual load impedance. We can see from the results for the micro-generator that the optimum load with the VM attached is approximately 16 times the optimum load without the VM. However the same theory would imply that for the macro generator the optimum load with the VM generator should be approximately 1.6 kΩ. However as show in figure 11 with a 1.6 kΩ load the transformation ratio of the VM circuit is approximately 1.5, thus indicating that the design of the VM does not match well the characteristics of the macro-generator. The micro generator is a better match than the macro generator with this VM circuit.

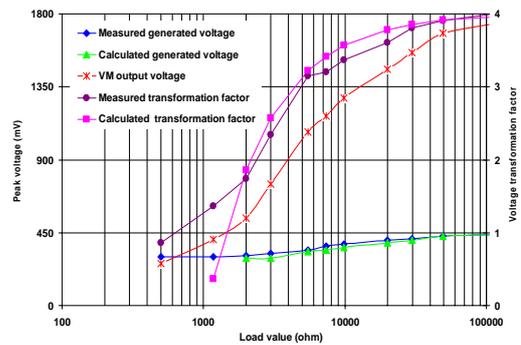

**Figure 11: Generated voltage and load voltage and the voltage ratio of macro generator with VM.**

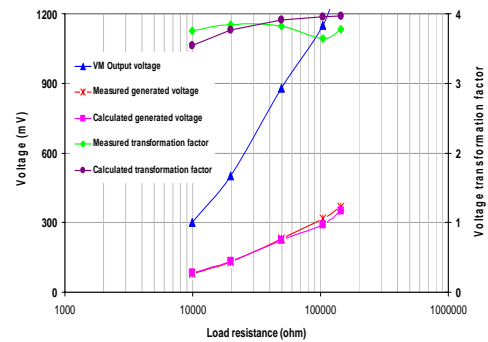

**Figure 12: Generated voltage and load voltage and the voltage ratio of micro generator with VM.**

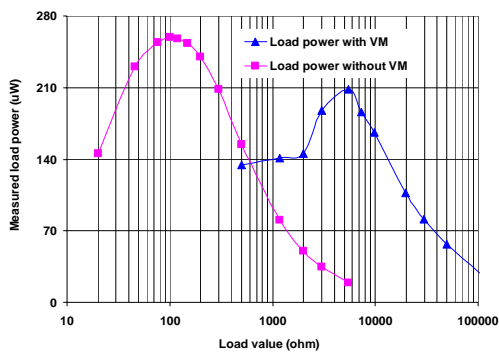

**Figure 13: Measured load power with and without voltage VM of macro generator.**

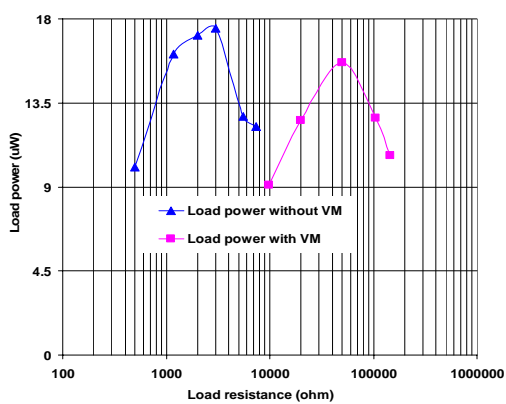

**Figure 14: Measured load power with and without VM of micro generator.**

### 7. CONCLUSIONS

A suitable voltage multiplier circuit for low voltage energy scavenger is introduced and characterized by signal generator and the real EM energy harvesting devices. The measured results showed 80% efficiency can be achieved at resonance frequency and the optimum load resistance with VM changes significantly at maximum power condition compare to resistive load.

### 7. ACKNOWLEDGEMENTS

The authors wish to acknowledge funding for this work under the European Union Framework 6 STREP project VIBES, project reference 507911 and the Higher Education Authority of Ireland fund for Digital Research.